\documentclass[preprint,showpacs,preprintnumbers,amsmath,amssymb]{revtex4}
\usepackage{epsfig}
\usepackage{graphicx}
\usepackage{dcolumn}
\usepackage{bm}

\newcommand{\ord}{{\cal O}}
\def\beq{\begin{equation}}
\def\eeq{\end{equation}}
\def\eeqn{\end{equation}}
\newcommand\iden{\leavevmode\hbox{\small1\normalsize\kern-.33em1}}

\newcommand{\bea} {\begin{eqnarray}}
\newcommand{\eea} {\end{eqnarray}}

\let\jnfont=\rm
\def\NPB#1,{{\jnfont Nucl.\ Phys.\ B }{\bf #1},}
\def\PLB#1,{{\jnfont Phys.\ Lett.\ B }{\bf #1},}
\def\EPJC#1,{{\jnfont Eur.\ Phys.\ Jour.\ C }{\bf #1},}
\def\PRD#1,{{\jnfont Phys.\ Rev.\ D }{\bf #1},}
\def\PRL#1,{{\jnfont Phys.\ Rev.\ Lett.\ }{\bf #1},}
\def\MPLA#1,{{\jnfont Mod.\ Phys.\ Lett.\ A }{\bf #1},}
\def\JPG#1,{{\jnfont J.\ Phys.\ G }{\bf #1},}
\def\CTP#1,{{\jnfont Commun.\ Theor.\ Phys.\ }{\bf #1},}
\def\JHEP#1,{{\jnfont JHEP \ }{\bf #1},}
\def\NPPS#1,{{\jnfont Nucl.\ Phys.\ Proc.\ Suppl.\ }{\bf #1},}
\def\CPC#1,{{\jnfont Computl.\ Phys.\ Commun.\ }{\bf #1},}
\def\CPL#1,{{\jnfont Chin.\ Phys.\ Lett. }{\bf #1},}
\def\AJS#1,{{\jnfont Astrophys.\ J.\ Suppl. }{\bf #1},}
\def\PR#1,{{\jnfont Phys.\ Rept. }{\bf #1},}
\def\AP#1,{{\jnfont Astropart.\ Phys. }{\bf #1},}
\def\EPL#1,{{\jnfont Europhys.\ Lett. }{\bf #1},}
\def\FP#1,{{\jnfont Fortsch.\ Phys. }{\bf #1},}
\def\JCAP#1,{{\jnfont JCAP \ }{\bf #1},}

\begin{document}

\title{\ \\[10mm] 130 GeV gamma-ray line and enhancement of $h\to \gamma\gamma$ in the Higgs triplet model plus a scalar dark matter}

\author{Lei Wang, Xiao-Fang Han}

\affiliation{ Department of Physics, Yantai University, Yantai
264005, China}


\begin{abstract}
With a discrete $Z_2$ symmetry being imposed, we introduce a real
singlet scalar $S$ to the Higgs triplet model with the motivation of
explaining the tentative evidence for a line spectral feature at
$E_\gamma$ = 130 GeV in the Fermi LAT data. The model can naturally
satisfy the experimental constraints of the dark matter relic
density and direct detection data from Xenon100. The doubly charged
and one charged scalars can enhance the annihilation cross section
of $SS\to\gamma\gamma$ via the one-loop contributions, and give the
negligible contributions to the relic density.  $<\sigma
v>_{SS\to\gamma\gamma}$ for $m_{S}=130$ GeV can reach
$\ord(1)\times10^{-27} cm^3 s^{-1}$ for the small charged scalars
masses and the coupling constant of larger than 1. Besides, this
model also predict a second photon peak at 114 GeV from the
annihilation $SS\to\gamma Z$, and the cross section is approximately
0.76 times that of $SS\to\gamma \gamma$, which is below the upper
limit reported by Fermi LAT. Finally, the light charged scalars can
enhance LHC diphoton Higgs rate, and make it to be consistent with
the experimental data reported by ATLAS and CMS.

\end{abstract}



\maketitle

\section{Introduction}

Recently, several groups \cite{12080009-1,12080009-2-3,12080009-4}
have reported a line spectral feature at $E_\gamma$ = 130 GeV in
publicly available data from the Fermi Large Area Telescope (LAT)
\cite{12080009-5}. Moreover, Ref. \cite{12080009-4,12080009-6}
reported the hints of a second line at around 111 GeV  with less
statistically significant. The sharp pick of the gamma-ray around
130 GeV can be explained by the 130 GeV dark matter (DM)
annihilating to two photons, whose cross section $<\sigma
v>_{SS\to\gamma\gamma}$ is
$1.27\pm0.32^{+0.18}_{-0.28}\times10^{-27}cm^3 s^{-1}$  ($
2.27\pm0.57^{+0.32}_{-0.51}\times10^{-27}cm^3 s^{-1}$) for Einasto
(NFW) DM profile employed \cite{12080009-1}. Besides, the line at
130 GeV can also be produced by the 142 GeV (155 GeV) DM
annihilating into $\gamma Z$ ($\gamma h$ with $h$ being a 125 GeV
Higgs boson). The Fermi LAT collaboration takes slightly different
search regions and methodology, and sets an upper limit of $<\sigma
v>_{SS\to\gamma\gamma} < 1.4\times10^{-27}cm^3 s^{-1}$, which is in
mild tension with the claimed signal \cite{12056811-26}.

The cross section of $SS\to\gamma\gamma$ ($1.27\times10^{-27}cm^3
s^{-1}$) required by the claimed 130 GeV gamma-ray line signal is
approximately 0.042 in units of the thermal relic density value,
$<\sigma v>_0 = 3\times10^{-26}cm^3 s^{-1}$ \cite{12074981-1}. Since
the DM is in general electrically neutral, $SS\to\gamma\gamma$
should arise at one-loop through the virtual massive charged
particles. If the charged particles at the loop are lighter than the
DM, the corresponding tree-level cross sections for annihilating to
these charged particles will exceed that of the loop-level process
to $\gamma\gamma$ by many orders of magnitude, which conflicts with
the total annihilation cross section to generate the observed relic
density. In addition, an enormous annihilation cross section to
charged particle is disfavored by the gamma-ray constraints from
observations of the Galactic Center and elsewhere
\cite{12056811-26,12080009-23}. A variety of DM models have been
proposed to solve this issue \cite{12052688,12056811,raymodel}. Ref.
\cite{12052688} shows that a multi-charged and colored scalar $X$
can enhance $<\sigma v>_{SS\to\gamma\gamma}$ to
$\ord{(1)}\times10^{-27}cm^3 s^{-1}$ via the interaction of
$\lambda_X SSXX$ at one-loop, and not lead to the conflict with the
relic density for its mass is larger than that of DM. In addition,
the LHC diphoton Higgs rate is also enhanced by the scalar.

To construct a DM model economically, a real singlet scalar is
respectively added to the standard model \cite{smd} and two Higgs
doublet model \cite{2hd} with a discrete $Z_2$ symmetry being
imposed. These models can satisfy naturally the constraints from the
DM relic density and direct detection data, but hardly accommodate
the claimed 130 GeV gamma-ray line signal \cite{smd-ph,2hd-ph}. In
this paper, we introduce such a scalar $S$ to the Higgs triplet
model (HTM) which contains a complex doublet Higgs field and a
complex triplet Higgs field with hypercharge $Y = 2$ \cite{htm}. In
the original HTM, several physical Higgs bosons remain after the
spontaneous symmetry breaking, including two CP-even ($h$ and $H$),
one CP-odd ($A$), one charged ($H^\pm$) and one doubly charged Higgs
scalars ($H^{\pm\pm}$). The charged scalars $H^{\pm\pm}$ and $H^\pm$
can enhance the cross section of $SS\to\gamma\gamma$ at one-loop.
Besides, the SM-like Higgs decay into two photon can be enhanced by
these charged scalars, which is favored by the new ATLAS and CMS
data. The new Higgs data has been discussed in the HTM
\cite{12060535,htmrr1,htmrr2}, the minimal supersymmetric standard
model (MSSM) \cite{11125453-26272829}, the next-to-MSSM
\cite{11125453-3537}, and other extensions of Higgs models
\cite{hrrmodel}.

This work is organized as follows. In Sec. II, we introduce a real
single scalar DM to the Higgs triplet model. In Sec. II, we study
the constraints of DM relic density and direct detection data. In
Sec. III, we calculate the cross sections of $<\sigma
v>_{SS\to\gamma\gamma}$ and $<\sigma v>_{SS\to\gamma Z}$. In Sec.
IV, we discuss the enhancement of LHC diphoton Higgs rate. Finally,
we give our conclusion in Sec. V.

\section{The Higgs triplet model plus a scalar DM (HTMD)}
In the HTM \cite{htm}, a complex $\rm{SU(2)_L}$ triplet scalar field
$\Delta$ with Y = 2 is added to the SM Lagrangian in addition to the
doublet field $\Phi$. These fields can be written as
\begin{eqnarray}
\Delta &=\left(
\begin{array}{cc}
\delta^+/\sqrt{2} & \delta^{++} \\
\delta^0 & -\delta^+/\sqrt{2}\\
\end{array}
\right),  \qquad \Phi=\left(
                    \begin{array}{c}
                      \phi^+ \\
                      \phi^0 \\
                    \end{array}
                  \right).
\end{eqnarray}
 The renormalizable scalar potential can
be written as \cite{htmpotent}
\begin{eqnarray}\label{potent}
V&=&-m_\Phi^2{\Phi^\dagger{\Phi}}+\frac{\lambda}{4}(\Phi^\dagger{\Phi})^2+
M_\Delta^2Tr(\Delta^{\dagger}{\Delta}) +
\lambda_1(\Phi^\dagger{\Phi})Tr(\Delta^{\dagger}{\Delta}) \\ & + &
\lambda_2(Tr\Delta^{\dagger}{\Delta})^2
+\lambda_3Tr(\Delta^{\dagger}{\Delta})^2 +
\lambda_4{\Phi^\dagger\Delta\Delta^{\dagger}\Phi}+
[\mu(\Phi^T{i}\tau_2\Delta^{\dagger}\Phi)+h.c.].\nonumber
\end{eqnarray}
The Higgs doublet and triplet fields can acquire vacuum expectation
values
\begin{equation}
\langle \Phi \rangle = \frac{1}{\sqrt{2}} \left(
                    \begin{array}{c}
                      0 \\
                      v_d \\
                    \end{array}
                  \right), \qquad \langle \Delta \rangle = \frac{1}{\sqrt{2}}
\left(
\begin{array}{cc}
0 & 0 \\
v_t & 0\\
\end{array}
\right)
\label{vacuum}
\end{equation} with $v_{SM}^2=v_d^2+4v_t^2\approx(246~\rm{GeV})^2$.

After the spontaneous symmetry breaking, the Lagrangian of Eq.
(\ref{potent}) predicts the seven physical Higgs bosons, including
two CP-even ($h$ and $H$), one CP-odd ($A$), one charged ($H^\pm$)
and one doubly charged Higgs scalars ($H^{\pm\pm}$). These mass
eigenstates are in general mixtures of the doublet and triplet
fields. The experimental value of the $\rho$ parameter requires
$v^2_t/v^2_d$ to be much smaller than unity at tree-level, which
gives a upper bound of $v_t <$ 8 GeV. \cite{12060535,ro}. For a very
small $v_t$, the mixing angle in the CP-even sector $\alpha$ and
charged Higgs sector $\beta$ are approximately, \beq
\sin\alpha\simeq 2v_t/v_d, ~~~~~\sin\beta \simeq \sqrt{2}v_t/v_d,
\label{angle}\eeq and the mixing of the doublet and triplet fields
is nearly absent. For this case, the seven Higgs masses can be
obtained from the Lagrangian of Eq. (\ref{potent})
\cite{12060535,htmrr1},
\begin{eqnarray}
m_{h}^2&\simeq&\frac{\lambda}{2}v^2_d, \nonumber\\
m_{H}^2&\simeq&M^2_\Delta + (\frac{\lambda_1}{2}
 +\frac{\lambda_4}{2}) v_d^2 + 3(\lambda_2 +\lambda_3) v^2_t, \, \nonumber\\
m_{A}^2&\simeq&M^2_\Delta + (\frac{\lambda_1}{2}
 +\frac{\lambda_4}{2}) v_d^2 + (\lambda_2 +\lambda_3) v^2_t, \, \nonumber\\
m_{H^\pm}^2&=& M^2_\Delta + (\frac{\lambda_1}{2}
 +\frac{\lambda_4}{4}) v_d^2 + (\lambda_2  + \sqrt 2
 \lambda_3) v^2_t, \,
\nonumber \\
m_{H^{\pm\pm}}^2&=&M^2_\Delta + \frac{\lambda_1}{2} v_d^2 +\lambda_2
v_t^2.   \label{mass}
\end{eqnarray}
In the following discussions, we always assume the value of $v_t$ is
very small. We take $h$ as the 125 GeV SM-like Higgs boson, which is
from the Higgs doublet field. $H$, $A$, $H^\pm$ and $H^{\pm\pm}$ are
heavier than $h$, which are from the Higgs triplet field. The $h$
field couplings to $f\bar{f}$, $WW$ and $ZZ$ equal to those of SM
nearly. In addition, the scalar potential terms in Eq.
(\ref{potent}) contain the SM-like Higgs boson couplings to the
charged scalars \cite{htmrr1}, \beq g_{h H^{++}H^{--}} \approx -
\lambda_1v_d,\qquad g_{h H^+H^-} \approx - (\lambda_{1} +
\frac{\lambda_{4}}{2}) v_d. \label{eq:gcalHHp} \eeq However, the
similar couplings for $H$ are suppressed by the factor $\sin\alpha$,
$v_t$ or $\sin\beta$. Thus, the $H$ production cross section at the
collider is very small, which satisfies the constraints of the
present Higgs data easily.

Now we introduce the renormalizable Lagrangian of the real single
scalar $S$, \beq\label{VD} {\cal L}_S=\frac{1}{2}\partial^\mu
S\partial_\mu
S-\frac{m_0^2}{2}SS-\frac{\kappa_1}{2}\Phi^\dagger{\Phi}SS- \kappa_2
Tr(\Delta^{\dagger}{\Delta})SS -\frac{\kappa_s}{4}S^4. \eeq The
linear and cubic terms of the scalar S are forbidden by the $Z_2$
symmetry $S\rightarrow -S$. $S$ has a vanishing vacuum expectation
value which ensures the DM candidate $S$ stable. $\kappa_s$ is the
coupling constant of the DM self-interaction, which does not give
the contributions to the DM annihilation and Higgs signal. In order
to explain the 130 GeV gamma-ray line signal, we take DM mass as 130
GeV, which determines the value of $m_0$ by the relation of
$m_S=(m_0^2+\frac{1}{2}\kappa_1 v_d^2+\kappa_2 v_t^2)^{1/2}$. The
total DM annihilation cross section mainly depends on the
$\kappa_1$, which determines the couplings $hSS$ and $hhSS$.
$\kappa_2$ determines the couplings $HSS$, $HHSS$, $AASS$, $H^{\pm}
H^{\mp} SS$ and $H^{\pm\pm} H^{\mp\mp}SS$, where the coupling $HSS$
is suppressed by $v_t$. The couplings $H^{\pm} H^{\mp} SS$ and
$H^{\pm\pm} H^{\mp\mp}SS$ give the important contributions to
$XX\to\gamma\gamma$ at one-loop.

For $v_t < 10^{-4}$ GeV, $H^{\pm\pm}\to \ell^\pm\ell^\pm$ is the
dominant decay mode of $H^{\pm\pm}$. Assuming $Br(H^{\pm\pm}\to
\ell^\pm\ell^\pm)=1$, CMS presents the low bound 383 GeV on
$m_{H^{\pm\pm}}$ from the searches for $H^{\pm\pm}\to
\ell^\pm\ell^\pm$ via $q\bar{q}\to H^{\pm\pm}H^{\mp\mp}$ and
$q\bar{q}\to H^{\pm\pm}H^{\mp}$ production processes
\cite{12072666}. However, $H^{\pm\pm}\to W^\pm W^\pm$ and
$H^{\pm\pm}\to H^\pm W^*$ are the dominant modes for $v_t > 10^{-4}$
GeV \cite{12060535,htmrr1,h2decay}, for which there have been no
direct searches. Therefore, the above bound on $m_{H^{\pm\pm}}$ can
not be applied to the case of $v_t > 10^{-4}$ GeV, and $H^{\pm\pm}$
could be much lighter in this scenario. In this paper, we take $v_t$
= 0.1 GeV and $m_{H^{\pm\pm}}$ to be as low as 140 GeV. LEP searches
for the charged scalar give the constraints on the possible
existence of light scalar \cite{lep}. A conservative lower bound on
$m_{H^{\pm}}$ should be larger than 100 GeV due to the absence of
non-SM events at LEP. To simplify the parameter space, we take the
triplet scalars to be degenerate, namely $\lambda_4$ = 0. We can
neglect the contributions of $\lambda_2$ and $\lambda_3$ to the
triplet scalars masses which are suppressed by $v_t^2$.

\section{dark matter relic density and direct detection}
\subsection{calculation of relic density}
The degenerate masses of the triplet scalars are taken to be larger
than 140 GeV. Thus, the annihilation processes $SS\to HH,~AA,~H^\pm
H^\mp, ~H^{\pm\pm} H^{\mp\mp}$ are forbidden for $m_S$ =130 GeV.
When the triplet scalars masses are slightly larger than DM mass,
the cross section of the forbidden annihilation channel is important
\cite{forbidden}. Here, we do not consider this scenario. Since the
$H$ field couplings to $SS$, $f\bar{f}$, $WW$ and $ZZ$ are
suppressed by $v_t$ or $\sin\alpha$, the s-channel annihilation
processes mediated by $H$ give a negligible contributions to total
DM annihilation cross section. Therefore, the main annihilation
processes include $SS\to f\bar{f}$, $SS\to WW$, $SS\to ZZ$ which
proceed via an s-channel $h$ exchange, and $SS\to hh$ which proceeds
via a 4-point contact interaction, an s-channel $h$ exchange and t-
and u-channel $S$ exchange. The total annihilation cross section
times the relative velocity $v$ for these processes is given as
\cite{10062518}, \beq \sigma
v=\sigma_{ff}v+\sigma_{WW}v+\sigma_{ZZ}v+\sigma_{hh}v,\eeq
\begin{eqnarray}
\sigma_{ff}v &=& \sum_f \frac{\kappa_1^2}{4\pi} \frac{
m_f^2}{(s-m_{h}^2)^2}
(1-\frac{4m_f^2}{s})^{3/2}, \label{FF} \nonumber\\
\sigma_{WW}v &=&  \frac{\kappa_1^2 }{8 \pi} \frac{s}{(s-m_{h}^2)^2}
\sqrt{1-\frac{4 m_{W}^2}{s}} \left(1- \frac{4m_{W}^2}{s}+ \frac{12
m_{W}^4}{s^2}
\right),\nonumber\\
\sigma_{ZZ}v &=&  \frac{\kappa_1^2 }{16 \pi} \frac{s}{(s-m_{h}^2)^2}
\sqrt{1-\frac{4 m_{Z}^2}{s}} \left(1- \frac{4m_{Z}^2}{s}+ \frac{12
m_{Z}^4}{s^2}
\right),\nonumber\\
\sigma_{hh}v &=&  \frac{\kappa_1^2 }{16 \pi s} \sqrt{1-\frac{4
m_{h}^2}{s}} \left[ \left(\frac{s+ 2 m_{h}^2}{s - m_{h}^2}\right)^2
- \frac{8 \kappa_1 v^2}{s-2 m_{h}^2} \frac{s+ 2 m_{h}^2}{s -
m_{h}^2} F(\xi) \right. \nonumber \\ & &+ \left. \frac{8 \kappa_1^2
v^4}{(s-2 m_{h}^2)^2} \left( \frac{1}{1-\xi^2} + F(\xi)\right)
\right]. \label{hh}
\end{eqnarray}
where $F(\xi)\equiv\mbox{arctanh}(\xi)/\xi$ with $\xi
\equiv\sqrt{(s-4m_{h}^2)(s-4m_D^2)}/(s-2m_{h}^2)$, and $s$ is the
squared center-of-mass energy.

The thermally averaged annihilation cross section times the relative
velocity, $<\sigma v>$, is well approximated by a non-relativistic
expansion, \beq <\sigma v> = a + b <v^2> +\ord(<v^4>) \simeq a+
6b\frac{T}{m_S}.\eeq  The freeze-out temperature $T_f$ is defined by
solving the following equation \cite{omiga}, \beq
x_f=\ln\frac{0.038gm_{pl}m_S<\sigma v>}{g_*^{1/2}x_f^{1/2}}.\eeq
Where $x_f=\frac{m_S}{T_f}$ and $m_{pl}=1.22\times10^{19}$ GeV.
$g_*$ is the total number of effectively relativistic degrees of
freedom at the time of freeze-out \cite{gstar}. $g=1$ is the
internal degrees of freedom for the scalar DM $S$. The present-day
abundance of $S$ is approximately \cite{omiga} \beq \Omega h^2\simeq
\frac{1.07\times
10^{9}}{m_{pl}}\frac{x_f}{\sqrt{g_*}}\frac{1}{(a+3b/x_f)}.\eeq

The relic density from the WMAP 7-year result \cite{omiga-ex} is
\beq \Omega_{DM} h^2= 0.1123 \pm 0.0035.\eeq

\subsection{Calculation of the spin-independent cross section between $S$ and nucleon}
 The results of DM-nucleus elastic scattering experiments are presented
in the form of a normalized DM-nucleon scattering cross section in
the spin-independent case. In the HTMD, the elastic scattering of
$S$ on a nucleon receives the dominant contributions from the $h$
exchange diagrams, which is given as \cite{sigis},
 \beq \sigma_{Sp(n)}^{SI}=\frac{m_{p(n)}^{2}}{4\pi\left(m_{S}+m_{p(n)}\right)^{2}}
    \left[f^{p(n)}\right]^{2},
\eeq where \beq
f^{p(n)}=\sum_{q=u,d,s}f_{T_{q}}^{p(n)}\mathcal{C}_{S
q}\frac{m_{p(n)}}{m_{q}}+\frac{2}{27}f_{T_{g}}^{p(n)}\sum_{q=c,b,t}\mathcal{C}_{S
q}\frac{m_{p(n)}}{m_{q}},\label{fpn} \eeq with $\mathcal{C}_{S
q}=\frac{\kappa_1 m_q}{m_h^2}$ \cite{twin51},
\begin{eqnarray}
f_{T_{u}}^{(p)}\approx0.020,\quad & f_{T_{d}}^{(p)}\approx0.026,\quad & f_{T_{s}}^{(p)}\approx0.118,\quad  f_{T_{g}}^{(p)}\approx0.836,\nonumber \\
f_{T_{u}}^{(n)}\approx0.014,\quad &
f_{T_{d}}^{(n)}\approx0.036,\quad & f_{T_{s}}^{(n)}\approx0.118,
\quad f_{T_{g}}^{(n)}\approx0.832. \label{eq:neuclon-form}
\end{eqnarray}
In fact, here $\sigma_{Sp}^{SI}\approx \sigma_{Sn}^{SI}$. The recent
data on direct DM search from Xenon100 put the most stringent
constraint on the cross section \cite{12076308-22}.

\subsection{results and discussions}
\begin{figure}[tb]
    \epsfig{file=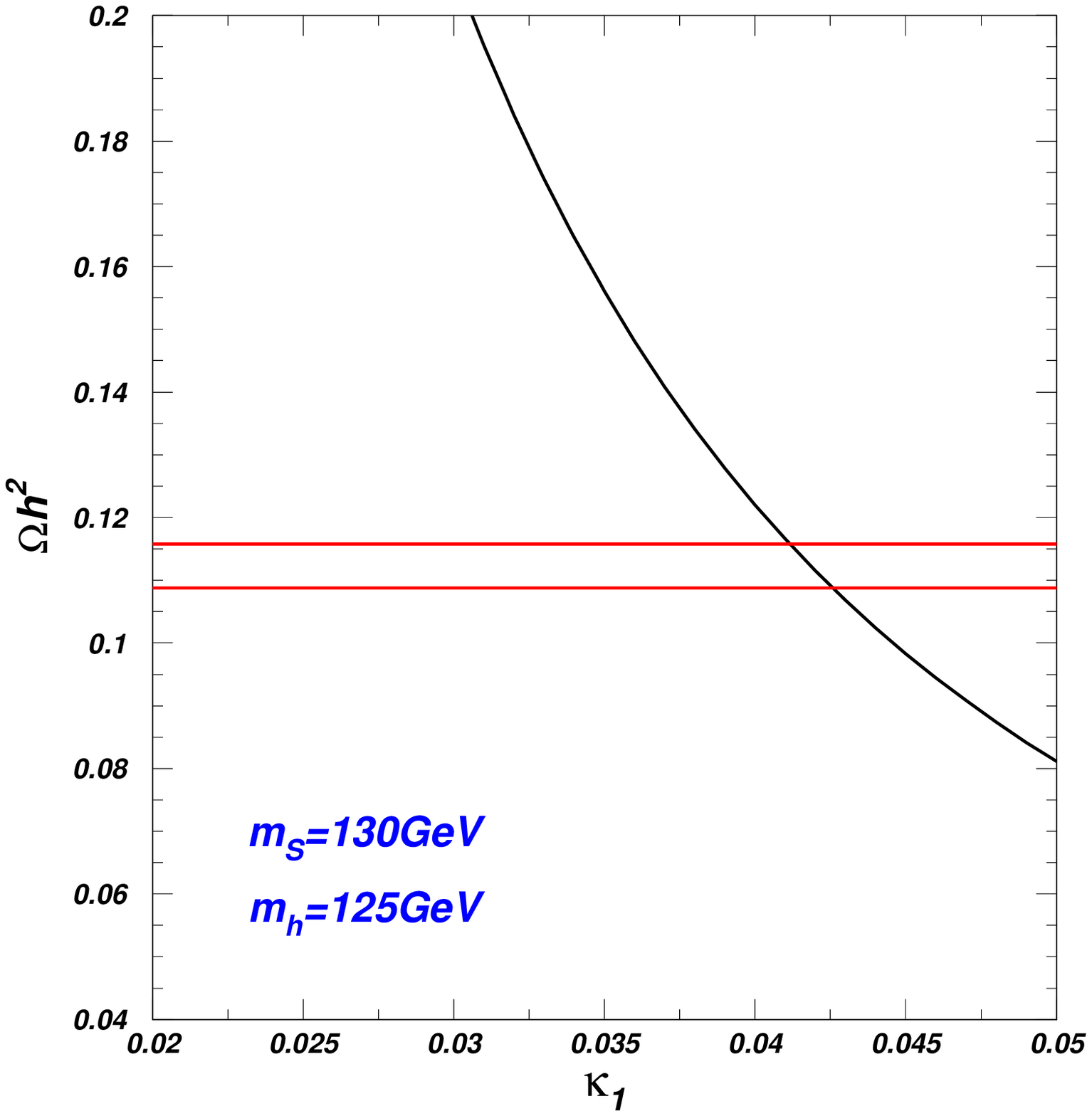,height=7cm}
    \epsfig{file=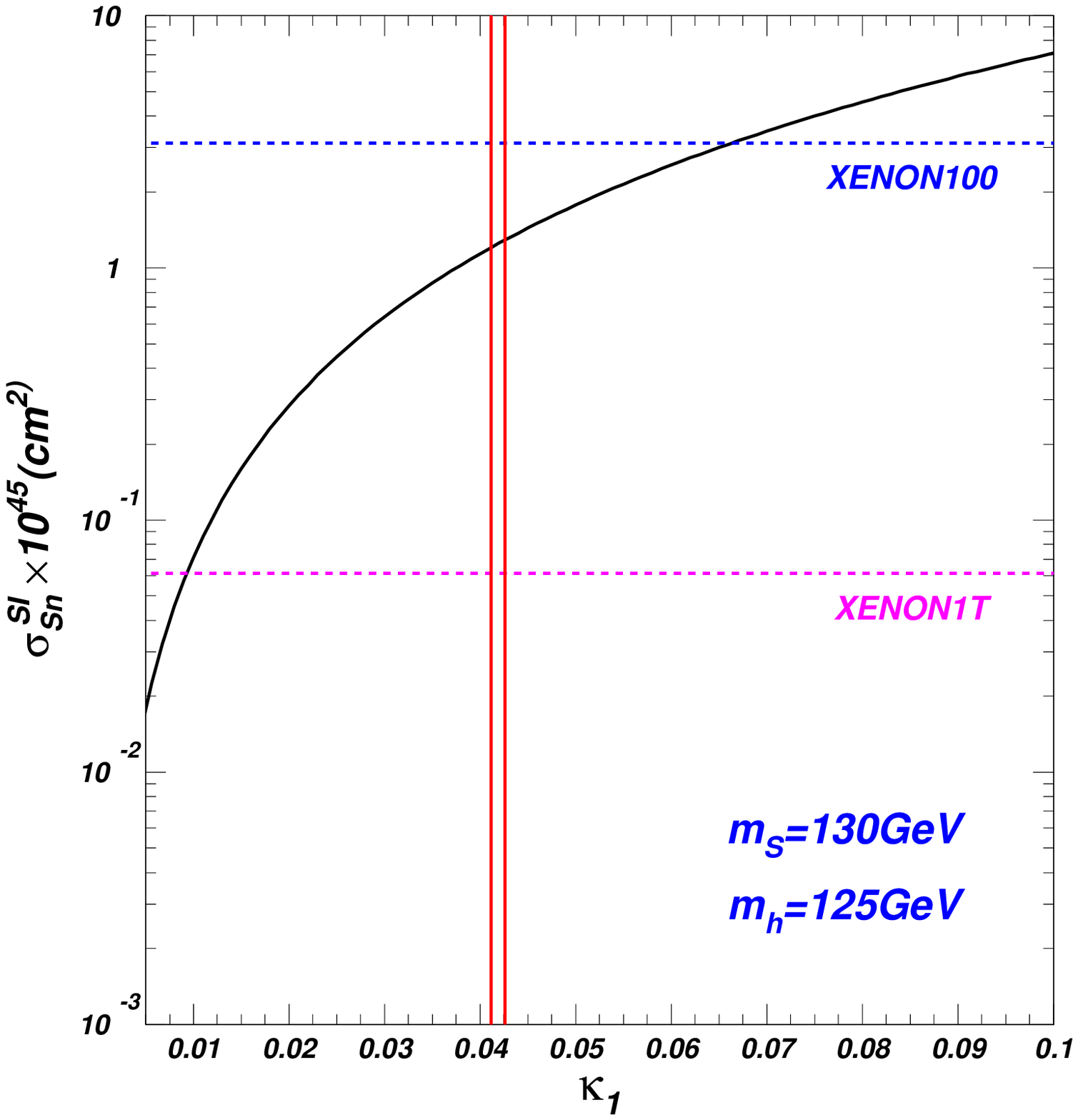,height=7cm}
\vspace{0.0cm} \caption{Left panel: The dark matter relic density
versus $\kappa_1$. The horizontal lines show the corresponding
bounds from experimental data of the WMAP 7-year. Right panel: the
spin-independent cross section between $S$ and the nucleon versus
 $\kappa_1$. The horizontal lines show the upper bound from Xenon100
 and the sensitivity of projected Xenon1T. The vertical lines
 show the range of $\kappa_1$ constrained by relic density.} \label{dencd}
\end{figure}

In our calculations, $m_S$ = 130 GeV and $m_h$ = 125 GeV are fixed.
Thus, both the relic density and the spin-independent cross section
between $S$ and the nucleon are only sensitive to the parameter
$\kappa_1$. In Fig. \ref{dencd}, we plot $\Omega h^2$ and
$\sigma_{Sn}^{SI}$ versus the $\kappa_1$, respectively. The left
panel of Fig. \ref{dencd} shows that $\kappa_1$ should be around
0.04 to get the correct DM relic abundance. For such value of
$\kappa_1$, the right panel shows that $\sigma_{Sn}^{SI}$ is around
$1.2\times 10^{-45} cm^2$, which is below the upper bound presented
by Xenon100 data and accessible at the future Xenon1T.

Refs. \cite{12056811-26,12080009-23,conti1} derive the limits on DM
annihilating to $f\bar{f}$ and $WW$ from the gamma-ray continuum, at
the level of $<\sigma
v>_{f\bar{f},WW}\lesssim\ord{(\rm{few})}\times10^{-25}cm^{-3}s^{-1}$,
depending on the final state particles. For $\kappa_1 =0.042$ which
is favored by the DM relic density, $<\sigma
v>_{e\bar{e}}\simeq1.3\times10^{-37}cm^{-3}s^{-1}$,  $<\sigma
v>_{\mu\bar{\mu}}\simeq5.5\times10^{-33}cm^{-3}s^{-1}$, $<\sigma
v>_{\tau\bar{\tau}}\simeq1.6\times10^{-30}cm^{-3}s^{-1}$, $<\sigma
v>_{b\bar{b}}\simeq2.6\times10^{-29}cm^{-3}s^{-1}$, and $<\sigma
v>_{WW}\simeq1.1\times10^{-26}cm^{-3}s^{-1}$, which satisfy easily
the limits of the continuum gamma-ray observations, respectively.

\section{gamma-ray lines from $SS\to\gamma\gamma$ and $SS\to \gamma Z$}

\subsection{130 GeV gamma-ray line from $SS\to\gamma\gamma$}

\begin{figure}[tb]
    \epsfig{file=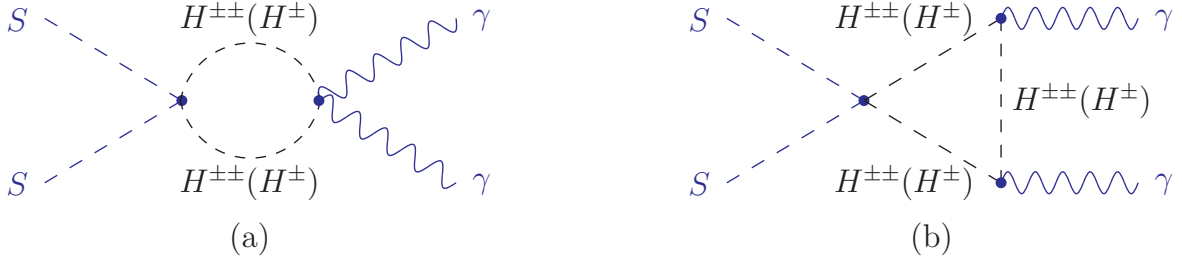,height=3.5cm}
\vspace{-0.4cm} \caption{The Feynman diagrams for
$SS\to\gamma\gamma$, which give the dominant contributions to the
annihilation process.} \label{fmt}
\end{figure}

The annihilation $SS\to\gamma\gamma$ may be radiatively induced by
massive charged particles in the loop. The charged scalars
$H^{\pm\pm}$ and $H^{\pm}$ can give the dominant contributions to
this annihilation process via the couplings $H^{\pm\pm}H^{\mp\mp}SS$
and $H^{\pm}H^{\mp}SS$, and the relevant Feynman diagrams are
depicted in Fig. \ref{fmt}. Besides, there is another type Feynman
diagram for $SS\to\gamma\gamma$ in which s-channel $h$ or $H$
exchange is combined with a charged particle loop. The contributions
of the diagram can be sizably enhanced for $m_h$ ($m_H$) $\sim$
2$m_S$ =260 GeV and the charged particle with an around 130 GeV mass
\cite{12056811}. For the SM-like Higgs $h$, its mass is 125 GeV and
the relic density requires $\kappa_1$ to be around 0.04, which
suppresses the coupling $hSS$. Although we may take $m_H$ =260 GeV,
the coupling $HSS$ is suppressed by $v_t$. Therefore, the
contributions from the type diagram are negligible compared to those
of Fig. \ref{fmt}. The annihilation cross section corresponding to
the diagrams of Fig. \ref{fmt} is approximately given by \beq
<\sigma v>_{SS\to \gamma\gamma}\simeq\frac{\alpha^2
\kappa_2^2}{32\pi^3 m_S^2}\Bigg| 4E(\tau_{H^{\pm\pm}})+
E(\tau_{H^{\pm}})\Bigg|^2 = \frac{25\alpha^2 \kappa_2^2}{32\pi^3
m_S^2}E(\tau_{H^{\pm\pm}})^2\eeq with
$\tau_{H^{\pm\pm}}=\frac{m_{H^{\pm\pm}}^2}{m_S^2}$,
$\tau_{H^\pm}=\frac{m_{H^\pm}^2}{m_S^2}$, and
$E(\tau)=1-\tau[\sin^{-1}(1/\sqrt{\tau})]^2$. For the second
equation, we take $m_{H^{\pm\pm}}=m_{H^{\pm}}$.

\begin{figure}[tb]
    \epsfig{file=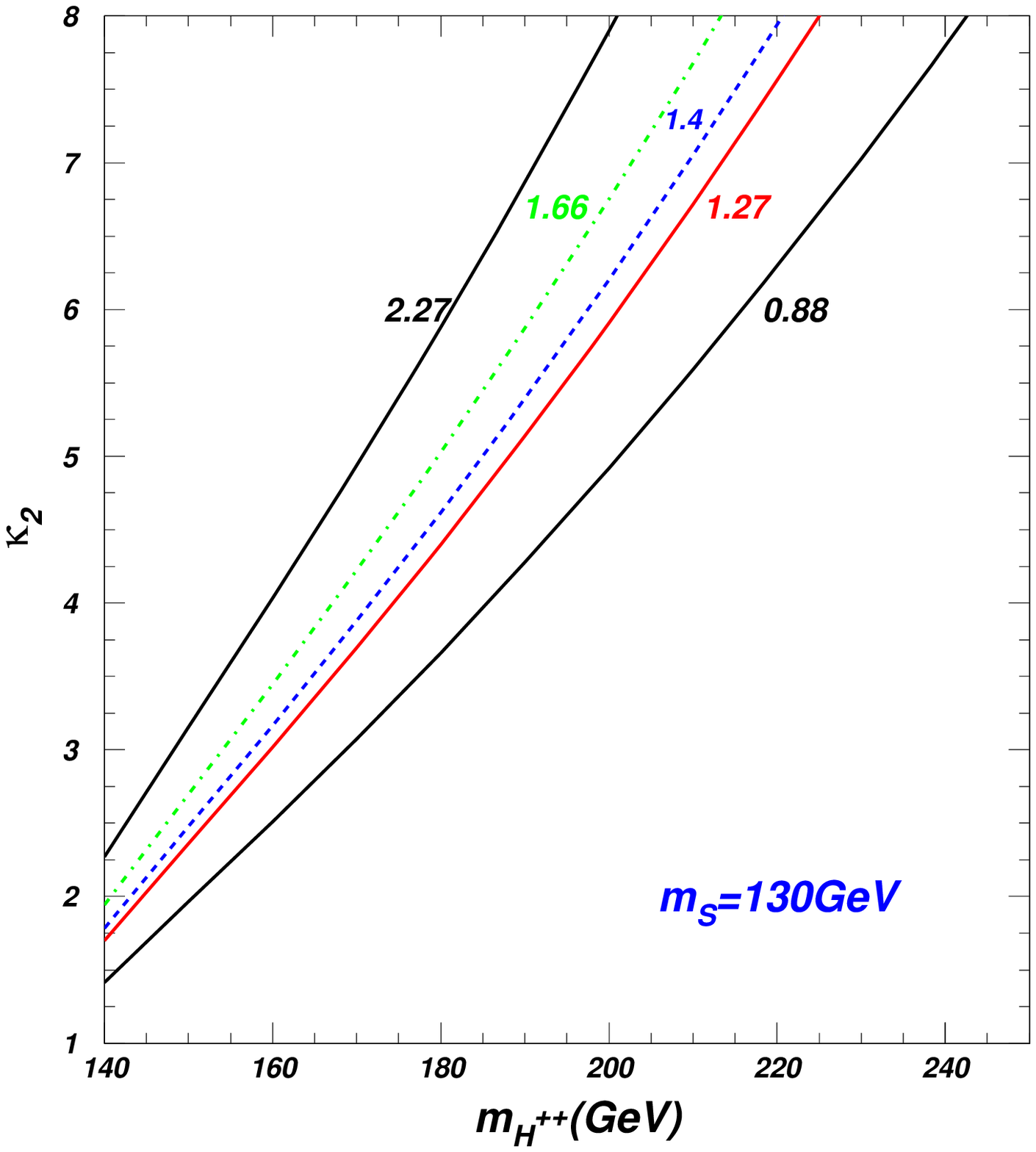,height=8.5cm}
\vspace{-0.4cm} \caption{The contours for $<\sigma v>_{
SS\to\gamma\gamma}$ in the plane of $\kappa_2$ versus
$m_{H^{\pm\pm}}$. The numbers on the cures denote $<\sigma v>_{
SS\to\gamma\gamma}/1.0\times10^{-27}cm^3s^{-1}$.}
\label{rayk2}\end{figure}

The $H^{\pm}$ and $H^{\pm\pm}$ contributions are constructive each
other. Because $H^{\pm\pm}$ has an electric charge of $\pm 2$, the
$H^{\pm\pm}$ contributions are enhanced by a relative factor 4 in
the amplitude. Fig. \ref{rayk2} shows some contours for $<\sigma
v>_{SS\to \gamma\gamma}=0.88\times 10^{-27}cm^3s^{-1},~1.27\times
10^{-27}cm^3s^{-1},~1.4\times 10^{-27}cm^3s^{-1},~1.66\times
10^{-27}cm^3s^{-1}$, and $2.27\times 10^{-27}cm^3s^{-1}$ in the
plane of $\kappa_2$ versus $m_{H^{\pm\pm}}$. From Fig. \ref{rayk2},
we find that, in order to obtain $<\sigma v>_{SS\to
\gamma\gamma}=1.27\times 10^{-27}cm^3s^{-1}$, the minimal value of
$\kappa_2$ should be from 1.7 to 4.0 for $m_{H^{\pm\pm}}$ in the
range of 140 GeV and 180 GeV. As the increasing of $m_{H^{\pm\pm}}$,
the corresponding $\kappa_2$ is required to increase, which will be
constrained by the perturbation of the theory.

\subsection{114 GeV gamma-ray line from $SS\to\gamma Z$}
We can obtain the Feynman diagrams of $SS\to \gamma Z$ and $SS\to
\gamma h$ by replacing a $\gamma$ with $Z$ and $h$ in the Fig.
\ref{fmt}, respectively. The cross section of $SS\to \gamma h$ is
zero due to the charge-conjugation invariance of the interactions
involved. The cross section of $SS\to \gamma Z$ is related to that
of $SS\to \gamma\gamma$, which is approximately given by \beq
\frac{<\sigma v>_{SS\to\gamma Z}}{<\sigma v>_{SS\to\gamma
\gamma}}\simeq
2(cot2\theta_W)^2(1-\frac{m_Z^2}{4m_S^2})^{1/2}=0.76.\eeq
 The energy of this single photon is given by $E_\gamma =
m_S(1-\frac{m_Z^2}{4m_S^2})=114 ~\rm{GeV}$. The current Fermi LAT
upper limit on $<\sigma v>_{SS\to\gamma Z}$ for $E_\gamma$ = 110 GeV
is $2.6\times 10^{-27} cm^3 s^{-1}$ ($3.6\times 10^{-27} cm^3
s^{-1}$ ) for Einasto (NFW) DM profile employed \cite{12056811-26}.
For $<\sigma v>_{SS\to\gamma \gamma}=1.27\times 10^{-27} cm^3
s^{-1}$, the prediction value of $<\sigma v>_{SS\to\gamma Z}$ is
below the upper bound presented by Fermi LAT.

\section{LHC diphoton Higgs rate}
To some extent, the decay $h\to \gamma\gamma$ is related to the
 annihilation process $SS\to \gamma\gamma$, since the doubly charged and one charged
scalars can contribute to both $SS\to \gamma\gamma$ and $h\to
\gamma\gamma$. It is necessary to restudy the LHC diphoton Higgs
rate although it has been studied in detail \cite{12060535,htmrr1}.

Since the new scalars and DM are heavy than the SM-like Higgs $h$,
$h$ does not have any new important decay modes compared to that of
SM. Except for the decay $h\to \gamma\gamma$, the other decay modes
and their widths are nearly the same both in HTMD and SM. The decay
width of $h\to \gamma\gamma$ is expressed as \cite{hrr1loop}
\begin{eqnarray}
\Gamma(h\to \gamma\gamma) = \frac{\alpha^2
  m^3_{h}}{256\pi^3v^2}\Bigg|  F_{1}(\tau_W)+ \sum_{i} N_{cf} Q^2_{f} F_{1/2}(\tau_f)
+  g_{_{H^{\pm}}}F_{0}(\tau_{H^\pm})+
4g_{_{H^{\pm\pm}}}F_{0}(\tau_{H^{\pm\pm}})\Bigg|^2 , \label{gamrr}
\end{eqnarray}
where \bea
&&\tau_W=\frac{4m_W^2}{m_h^2},~~~\tau_f=\frac{4m_f^2}{m_h^2},~~~
\tau_{H^\pm}=\frac{4m_{H^\pm}^2}{m_h^2},~~~\tau_{H^{\pm\pm}}=\frac{4m_{H^{\pm\pm}}^2}{m_h^2},\nonumber\\
&&g_{_{H^{\pm}}}=-\frac{v}{2m_{H^\pm}^2}g_{hH^+ H^-},~~~
g_{_{H^{\pm\pm}}}=-\frac{v}{2m_{H^{\pm\pm}}^2}g_{hH^{++}
H^{--}}.\eea $N_{cf}$, $Q_f$ are the color factor and the electric
charge respectively for fermion $f$ running in the loop. The
dimensionless loop factors for particles of spin given in the
subscript are:
\begin{eqnarray}
F_1 = 2+3\tau + 3\tau(2-\tau)f(\tau), \quad F_{1/2} =
-2\tau[1+(1-\tau)f(\tau)], \quad F_0 = \tau[1-\tau f(\tau)],
\label{ffun}\end{eqnarray} with
\begin{equation}
f(\tau) = \left\{ \begin{array}{lr}
[\sin^{-1}(1/\sqrt{\tau})]^2, & \tau \geq 1 \\
-\frac{1}{4} [\ln(\eta_+/\eta_-) - i \pi]^2, & \, \tau < 1
\end{array}  \right.
\end{equation}
where $\eta_{\pm} = 1 \pm \sqrt{1-\tau}$.

The Higgs boson production cross sections at the LHC are the same
both in the HTMD and SM. Therefore, the LHC diphoton rate of Higgs
boson in the HTMD normalized to the SM prediction can be written as
\beq
R_{\gamma\gamma}=\frac{Br(h\to\gamma\gamma)}{Br(h\to\gamma\gamma)^{SM}}
\simeq\frac{\Gamma(h\to\gamma\gamma)}{\Gamma(h\to\gamma\gamma)^{SM}}.\eeq
The new data of LHC presents the constraints on $R_{\gamma\gamma}$,
$R_{\gamma\gamma}=1.56\pm0.43$ for $m_h \simeq$ 125 GeV from CMS
\cite{CMS} and $R_{\gamma\gamma}=1.9\pm0.5$ for $m_h \simeq$ 126 GeV
from ATLAS \cite{ATLAS}.

\begin{figure}[tb]
    \epsfig{file=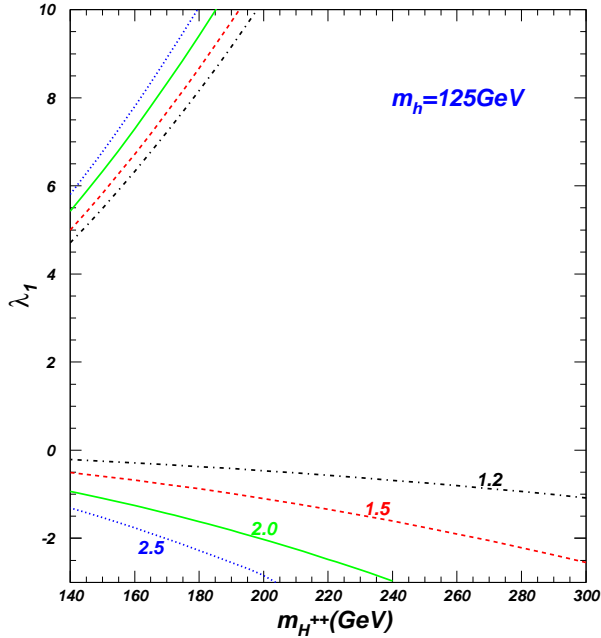,height=8.5cm}
\vspace{-0.4cm} \caption{The contours for $R_{\gamma\gamma}$ in the
plane of $\lambda_1$ versus $m_{H^{\pm\pm}}$. The numbers on the
cures denote the values of $R_{\gamma\gamma}$.} \label{rrk2}
\end{figure}

By tuning the values of $\lambda_2$ and $\lambda_3$, $-3 \leq
\lambda_1 \leq 10$ is allowed by the perturbative unitarity and
stability of the potential \cite{12060535}. Since the effects of
$\lambda_2$ and $\lambda_3$ on $R_{\gamma\gamma}$ are suppressed by
$v_t$, $R_{\gamma\gamma}$ is not sensitive to the choices of
$\lambda_2$ and $\lambda_3$. We take $\lambda_4$ = 0, which leads
that $H^{\pm\pm}$ and $H^{\pm}$ have the same masses, and their
couplings to $h$ are equal and proportional to $\lambda_1$.
Therefore, $R_{\gamma\gamma}$ is only sensitive to $m_{H^{\pm\pm}}$
and $\lambda_1$. Fig. \ref{rrk2} shows some contours for
$R_{\gamma\gamma}=1.2,~1.5,~2.0,~2.5$ in the plane of $\lambda_1$
versus $m_{H^{\pm\pm}}$. The $H^{\pm}$ and $H^{\pm\pm}$
contributions are constructive with those of $W$ boson for
$\lambda_1 <$  0, but destructively for $\lambda_1 >$ 0. From Fig.
\ref{rrk2}, we can find that, if $\lambda_1$ is larger than 0, 1.2
$< R_{\gamma\gamma} < $ 2.5 requires $\lambda_1 > $ 4 and
$m_{H^{\pm\pm}} < 200$ GeV, which is similar to that of $SS\to
\gamma\gamma$, namely a large coupling constant and the light
charged scalars. For $\lambda_1 <$ 0, the charged scalars masses can
be as high as 300 GeV.

\section{Conclusion}
In the framework of Higgs triplet model, a real single scalar $S$ is
introduced with a discrete $Z_2$ symmetry being imposed, which plays
the role of the DM candidate. The interaction between DM and SM-like
Higgs $h$ gives the dominant contributions to the total DM
annihilation cross section and cross section of between DM and
nucleon, which can make the model to satisfy the experimental
constraints of DM relic density and direct detection data from
Xenon100. The doubly charged scalar and one charged scalar can give
the important contributions to the annihilation process
$SS\to\gamma\gamma$ and the decay $h\to \gamma\gamma$. For these
charged scalars masses are suitable small, $<\sigma
v>_{SS\to\gamma\gamma}$ can be enhanced to
$\ord(1)\times10^{-27}cm^3 s^{-1}$, which gives a valid explanation
for the claimed 130 GeV gamma-ray line signal. The LHC diphoton rate
can be enhanced by a factor 1.2 $\sim$ 2.5, which fits the ATLAS and
CMS data well. Besides, the model also predicts a second 114 GeV
gamma-ray line from the $SS\to \gamma Z$ annihilation, whose cross
section is below the upper bound reported by Fermi LAT.

\section*{Acknowledgment}
We thank Wen Long Sang, Wenyu Wang and Jin Min Yang for discussions.
This work was supported by the National Natural Science Foundation
of China (NNSFC) under grant Nos. 11005089 and 11105116.


\begin{thebibliography}{99}
\bibitem{12080009-1} C. Weniger, \JCAP1208, 007 (2012) [arXiv:1204.2797].

\bibitem{12080009-2-3} T. Bringmann, X. Huang, A. Ibarra,
 S. Vogl, and C. Weniger, \JCAP1207, 054 (2012) [arXiv:1203.1312];
 E. Tempel, A. Hektor, and M. Raidal, arXiv:1205.1045.

\bibitem{12080009-4} M. Su and D. P. Finkbeiner, arXiv:1206.1616.

\bibitem{12080009-5} W. Atwood et al. (LAT Collaboration),
Astrophys. J. 697, 1071 (2009) [arXiv:0902.1089].

\bibitem{12080009-6} M. Su and D. P. Finkbeiner, arXiv:1207.7060.

\bibitem{12056811-26} M. Ackermann et al. [Fermi-LAT Collaboration],
 \PRD86, 022002 (2012) [arXiv:1205.2739].

\bibitem{12074981-1} G. Jungman, M. Kamionkowski and K. Griest, \PR267, 195 (1996)
 [hep-ph/9506380]; G. Bertone, D. Hooper and J. Silk, \PR405, 279 (2005)
 [hep-ph/0404175].


\bibitem{12080009-23}  M. Ackermann et al. (Fermi-LAT collaboration), \PRL107, 241302 (2011)
[arXiv:1108.3546].

\bibitem{12052688} J. M. Cline, \PRD86, 015016 (2012) [arXiv:1205.2688].

\bibitem{12056811} M. R. Buckley and D. Hooper, \PRD86, 043524 (2012)
[arXiv:1205.6811].

\bibitem{raymodel} A. Ibarra, S. L. Gehler and M. Pato, \JCAP1207, 043 (2012)
[arXiv:1205.0007]; E. Dudas, Y. Mambrini, S. Pokorski and A.
Romagnoni, arXiv:1205.1520;  K. Y. Choi and O. Seto, \PRD86, 043515
(2012) [arXiv:1205.3276]; H. M. Lee, M. Park and W. -I. Park,
arXiv:1205.4675; A. Rajaraman, T. M. P. Tait and D. Whiteson,
arXiv:1205.4723; B. S. Acharya, G. Kane, P. Kumar, R. Lu and B.
Zheng, arXiv:1205.5789; X. Chu, T. Hambye, T. Scarna and M. H. G.
Tytgat, arXiv:1206.2279; D. Das, U. Ellwanger and P. Mitropoulos,
\JCAP1208, 003 (2012) [arXiv:1206.2639]; Z. Kang, T. Li, J. Li and
Y. Liu, arXiv:1206.2863; I. Oda, arXiv:1207.1537; R. -Z. Yang, Q.
Yuan, L. Feng, Y. -Z. Fan and J. Chang, arXiv:1207.1621; L. Feng, Q.
Yuan, Y.-Z, Fan, arXiv:1206.4758; B. Yang, J. Shelton,
arXiv:1208.4100; L. Bergstrom, arXiv:1208.6028; J. M. Cline, A. R.
Frey, G. D. Moore, arXiv:1208.2685; S. Tulin, H.-B. Yu, K. M. Zurek,
arXiv:1208.0009; X.-Y. Huang, Q. Yuan, P.-F. Yin, X.-J. Bi, X.-L.
Chen, arXiv:1208.0267; R. Laha, K. C. Yu Ng, B. Dasgupta, S,
Horiuchi, arXiv:1208.5488; J. J. Fan, M. Reece, arXiv:1209.1097;
J.-C. Park, S. C. Park, arXiv:1207.4981.

\bibitem{smd} V. Silveira, A. Zee, \PLB161, (1985) 136.

\bibitem{2hd} C. Bird, R. Kowalewski, M. Pospelov, \MPLA 21, 457 (2006)
[arXiv:hep-ph/0601090].

\bibitem{smd-ph} K. Cheung, Y.-L. S. Tsai, P. Tseng, T. Yuan,
A. Zee, arXiv:1207.4930.

\bibitem{2hd-ph} X.-G. He, B. Ren, J. Tandean, \PRD85, 093019
(2012).

\bibitem{htm} W. Konetschny, W. Kummer, \PLB70, 433 (1977);
J. Schechter, J. W. F. Valle, \PRD 22, 2227 (1980); T. P. Cheng, L.
F. Li, \PRD22, 2860 (1980).


\bibitem{12060535} A. G. Akeroyd, S. Moretti, \PRD86, 035015 (2012) [arXiv:1206.0535].


\bibitem{htmrr1} A. Arhrib, R. Benbrik, M. Chabab, G. Moultaka, L.
Rahili, \JHEP1204, 136 (2012).

\bibitem{htmrr2} L. Wang, X.-F Han, arXiv:1206.1673.


\bibitem{11125453-26272829} M. Carena, S. Gori, N. R. Shah, and C.
E. M. Wagner, \JHEP1203, 014 (2012); J. Cao, Z. Heng, D. Li, J. M.
Yang, \PLB710, 665-670 (2012); \JHEP1203, 086 (2012); N.
Christensen, T. Han, S. Su, \PRD85,  115018 (2012)
[arXiv:1203.3207]; L. Roszkowski, E. M. Sessolo, Y.-L. Sming Tsai,
arXiv:1202.1503.

\bibitem{11125453-3537} J. F. Gunion, Y. Jiang, S. Kraml, \PLB710, 454-459 (2012);
 U. Ellwanger, \PLB698, 293 (2011); J. Cao, Z. Heng, T. Liu, J. M. Yang, \PLB703, 462
(2011); D. A. Vasquez, G. Belanger, C. Boehm, J. Da Silva, P.
Richardson, C. Wymant, \PRD86, 035023 (2012) [arXiv:1203.3446]; F.
King, M. Muhlleitner, R. Nevzorov, \NPB860, 207-244 (2012); J. Cao,
Z. Heng, J. M. Yang, J. Zhu, arXiv:1207.3698; J. Ke, M.-X. Luo,
L.-Y. Shan, K. Wang, L. Wang, arXiv:1207.0990.

\bibitem{hrrmodel}
T. Li, X. Wan, Y.-K. Wang, S.-H. Zhu, arXiv:1203.5083; L. Wang, J.
M. Yang, \PRD84, 075024 (2011);  A. Arhrib, R. Benbrik, N. Gaur,
\PRD85, 095021 (2012) [arXiv:1201.2644]; L. Wang, X.-F. Han,
\JHEP1205, 088 (2012); Y. Cai, W. Chao, S. Yang, arXiv:1208.3949; N.
Chen, H.-J. He, \JHEP1204, 062 (2012); M. R. Buckley, D. Hooper,
arXiv:1207.1445; P. Giardino, K. Kannike, M. Raidal, A. Strumia,
arXiv:1207.1347; S. Chang, C. A. Newby, N. Raj, C. Wanotayaroj,
arXiv:1207.0493; T. Abe, N. Chen, H.-J. He, arXiv:1207.4103.


\bibitem{htmpotent} E. Ma, M. Raidal and U. Sarkar, \PRL85, 3769
(2000); E. Ma, M. Raidal and U. Sarkar, \NPB615, 313 (2001); E. J.
Chun, K. Y. Lee and S. C. Park, \PLB566, 142 (2003).

\bibitem{ro} S. Kanemura, K. Yagyu, \PRD85, 115009 (2012) [arXiv:1201.6287].

\bibitem{12072666} CMS Collaboration, arXiv:1207.2666.

\bibitem{h2decay} A. G. Akeroyd and M. Aoki, \PRD72, 035011 (2005).

\bibitem{lep} DELPHI Collaboration, J. Abdallah et al., \EPJC34, 399
(2004).

\bibitem{forbidden} K. Griest, D. Seckel, \PRD43, 3191-3203 (1991).

\bibitem{10062518} W.-L. Guo, Y.-L. Wu, \JHEP1010, 083 (2010);
R. N. Lerner, J. McDonald, \PRD80, 123507 (2009); P. Burgess, M.
Pospelov, T. Veldhuis, \NPB619, 709-728 (2001).


\bibitem{omiga} E. Kolb and M. Turner, The Early Universe (Frontiers in Physics)
(Westview Press, 1994).

\bibitem{gstar} P. Gondolo and G. Gelmini, \NPB360, 145 (1991).

\bibitem{omiga-ex} E. Komatsu et al. [WMAP Collaboration], \AJS192,
18 (2011).


\bibitem{sigis} G. Jungman, M. Kamionkowski, K. Griest, \PR267, 195
(1996); M. A. Shifman, A. I. Vainshtein and V. I. Zakharov, \PLB78,
443 (1978).


\bibitem{twin51} J. R. Ellis, K. A. Olive, Y. Santoso and V. C. Spanos,
\PRD71, 095007 (2005) [hep-ph/0502001].

\bibitem{12076308-22} E. Aprile et al. [XENON100 Collaboration],
\PRL107, 131302 (2011) [arXiv:1104.2549].


\bibitem{conti1} W. Buchm$\ddot{u}$ller, M. Garny, arXiv:1206.7056;
T. Cohen, M. Lisanti, T. R. Slatyer and J. G. Wacker,
arXiv:1207.0800; I. Cholis, M. Tavakoli and P. Ullio,
arXiv:1207.1468.

\bibitem{hrr1loop} A. Djouadi, \PR459, 1 (2008).


\bibitem{CMS} S. Chatrchyan et al. [CMS Collaboration],
\PLB716, 30-61 (2012) [arXiv:1207.7235].


\bibitem{ATLAS} G. Aad et al. [ATLAS Collaboration], \PLB716, 1-29
(2012) [arXiv:1207.7214].


\end{thebibliography}
\end{document}